\documentclass[10pt]{iopart}
\usepackage{microtype}
\usepackage{iopams}
\usepackage{graphics}
\usepackage{cite}

\newcommand{\p}{\partial}
\newcommand{\be}{\begin{equation}}
\newcommand{\ee}{\end{equation}}
\newcommand{\tP}{\tilde{P}}
\newcommand{\tC}{\tilde{C}}

\begin{document}
\title[Power spectra Brownian motion with solid friction]{Exact power spectra of Brownian motion with solid friction}

\author{Hugo Touchette, Thomas Prellberg, and Wolfram Just}

\address{School of Mathematical Sciences,
Queen Mary University of London, London E1 4NS, UK}

\eads{
\mailto{h.touchette@qmul.ac.uk},
\mailto{t.prellberg@qmul.ac.uk},
\mailto{w.just@qmul.ac.uk}
}

\begin{abstract}
We study a Langevin equation describing the Brownian motion of an object subjected to a viscous drag, an external constant force, and a solid friction force of the Coulomb type. In a previous work [H.~Touchette, E.~Van der Straeten, W.~Just, \textit{J.~Phys.~A: Math.~Theor.}~\textbf{43}, 445002, 2010], we have presented the exact solution of the velocity propagator of this equation based on a spectral decomposition of the corresponding Fokker-Planck equation. Here, we present an alternative, exact solution based on the Laplace transform of this equation, which has the advantage of being expressed in closed form. From this solution, we also obtain closed-form expressions for the Laplace transform of the velocity autocorrelation function and for the power spectrum, i.e., the Fourier transform of the autocorrelation function. The behavior of the power spectrum as a function of the dry friction force and external forcing shows a clear crossover between stick and slip regimes known to occur in the presence of solid friction.
\end{abstract}


\section{Introduction}

We continue in this article our study of Brownian motion involving solid (dry or Coulomb) friction in addition to viscous friction; see \cite{baule2010,baule2011,touchette2010c}. As in these works, we consider the piecewise linear Langevin equation
\be
\dot{v}=-\gamma v - \Delta\sigma(v) +F + \sqrt{\Gamma}\, \xi(t),
\label{ab}
\ee
where $\gamma>0$ denotes the viscous coefficient, $\Delta>0$ the dry friction coefficient, $F$ an external constant forcing, $\xi(t)$ a Gaussian white noise and $\Gamma$ the related diffusion constant. The term $-\Delta\sigma(v)$, where $\sigma(v)$ denotes the sign of $v$ with the convention $\sigma(0)=0$, represents the dry friction force. Its physical interpretation follows by considering Eq.~(\ref{ab}) in the deterministic limit $\Gamma=0$: for $|F|<\Delta$, the stationary state of this equation is the ``sticking'' state $v=0$, whereas for $|F|>\Delta$, the stationary state is a ``sliding'' state with $v\neq 0$. Thus, to induce motion ($v\neq 0$) from rest ($v=0$), $F$ has to be larger than $\Delta$, the dry friction contact force.\footnote{Mathematically speaking, Eq.~(\ref{ab}) is incomplete: in order to ensure the existence of well-defined global solutions of this equation, we must ask in addition that all external forces vanish in the stick state $v = 0$ when $|F|\leq\Delta$.} 

The effect of dry friction on the properties of Brownian motion was studied by De Gennes \cite{gennes2005}, who showed, for the special case $\gamma=F=0$, that the velocity-velocity correlation function $\langle v(t) v(0)\rangle$ acquires a dependence on the noise power when $\Delta>0$. In \cite{touchette2010c}, we extended his study by obtaining eigenfunction expansions of the propagator $p(v,t|v_0,0)$ for the general case  $\gamma> 0$ and $F> 0$. From these expansions, which involve a special function known as the parabolic cylinder function, we were also able to compute $\langle v(t) v(0)\rangle$. Our main finding was that the stick and slip states of the deterministic system ($\Gamma=0$) translate into stick and slip regimes in the noisy system ($\Gamma\neq 0$), which are characterized by a strong and weak dependence, respectively, of the correlation time of the velocity-velocity correlation function with the external force $F$.

Our goal in this paper is to show that our exact results of \cite{touchette2010c} can be expressed in a more convenient way by solving the Fokker-Planck equation associated with Eq.~(\ref{ab}) in Laplace space rather than in direct space. The resulting expressions for the propagator and velocity-velocity correlation function are indeed somewhat more compact and more elegant than the eigenfunction expansions presented earlier. Our solution in Laplace space also enables us to complete the study of Eq.~(\ref{ab}) by deriving exact, closed-form expressions for the power spectrum of this equation, i.e., the Fourier transform of the velocity-velocity correlation function. These expressions extend early articles by Caughey and Dienes \cite{caughey1961} and by Atkinson and Caughey \cite{atkinson1968,atkinson1968b}, recently brought to our attention, which considered the power spectrum of the Langevin equation with pure Coulomb friction, i.e., the same case ($\gamma=F=0$) considered by De Gennes.

\section{Propagator}
\label{sec2}

As in \cite{touchette2010c}, we study the propagator $P(x,t|x',0)$ of Eq.~(\ref{ab}), expressed in terms of the non-dimensional variables $\sqrt{2\gamma/\Gamma} v \rightarrow x$, $\gamma t \rightarrow t$. The Fokker-Planck equation governing the evolution of this propagator has the form
\be
\frac{\p P}{\p t} = \frac{\p}{\p x} (x+\delta \sigma(x)-f) P+ \frac{\p^2 P}{\p x^2},
\label{aa}
\ee
where $\delta = \Delta\sqrt{2/(\gamma \Gamma)}$ measures the magnitude of the dry friction relative to the viscous damping while $f=F/\sqrt{2/(\gamma \Gamma)}$ stands for the external constant force, measured also against the viscous damping.

The time-independent or stationary solution of the Fokker-Planck equation has the standard form
\be
\rho_f(x)=\frac{e^{-\Phi(x)}}{Z}
\ee
in terms of the potential
\be
\Phi(x)=\frac{(|x|+\delta)^2}{2}-f x.
\ee 
Depending on the sign of the parameter $\delta$, the potential refers to a dry friction problem ($\delta>0$) or a Kramer-type tunneling problem ($\delta<0$).\footnote{Although we do not study the case $\delta<0$, all results derived here are also valid for this case.}

For piecewise-linear Fokker-Planck equations, the propagator can be obtained exactly, as was observed in \cite{atkinson1968,atkinson1968b}, by considering its Laplace transform:
\be
\tP(x,s|x',0)=\int_0^\infty e^{-s t} P(x,t|x',0) \, dt.
\label{ba}
\ee
With this transform, Eq.~(\ref{aa}) becomes a second-order ordinary differential equation 
\be
\fl s \tP(x,s|x',0) -\delta(x-x') = \frac{d}{d x} (x+\delta \sigma(x)-f) \tP(x,s|x',0)+ \frac{d^2 \tP}{d x^2},
\label{bb}
\ee
which can be solved in terms of parabolic cylinder functions. The detail of this solution is given in \ref{appa}. The final result obtained for positive values of the initial condition, i.e., $x'>0$, has for expression
\begin{eqnarray}
\tP(x,s|x',0) &=& g_<(s,\delta,f)\, \frac{\Gamma(s)}{\sqrt{2\pi}}\, e^{(x'+\delta-f)^2/4}\, D_{-s}(x'+\delta-f)\nonumber \\
& & \qquad \times e^{-(x-\delta-f)^2/4}\, D_{-s}(-x+\delta+f) 
\label{bca}
\end{eqnarray}
for $x<0$,
\begin{eqnarray}
\tP(x,s|x',0) &=& g_>(s,\delta,f) \, \frac{\Gamma(s)}{\sqrt{2\pi}} \, e^{(x'+\delta-f)^2/4}\, D_{-s}(x'+\delta-f)\nonumber \\
& & \qquad \times e^{-(x+\delta-f)^2/4}\, D_{-s}(x+\delta-f)\nonumber \\
& & +\frac{\Gamma(s)}{\sqrt{2\pi}}\, e^{(x'+\delta-f)^2/4}\, D_{-s}(x'+\delta-f)\nonumber \\
& & \qquad \times e^{-(x+\delta-f)^2/4}\, D_{-s}(-x-\delta+f) 
\label{bcb}
\end{eqnarray}
for $0<x<x'$, and
\begin{eqnarray}
\tP(x,s|x',0) &=& g_>(s,\delta,f) \, \frac{\Gamma(s)}{\sqrt{2\pi}}\, e^{(x'+\delta-f)^2/4}\, D_{-s}(x'+\delta-f) \nonumber \\
& & \qquad \times e^{-(x+\delta-f)^2/4} D_{-s}(x+\delta-f)\nonumber \\
& &+\frac{\Gamma(s)}{\sqrt{2\pi}}\, e^{(x'+\delta-f)^2/4}\, D_{-s}(-x'-\delta+f)\nonumber \\
& & \qquad \times e^{-(x+\delta-f)^2/4}\, D_{-s}(x+\delta-f) 
\label{bcc}
\end{eqnarray}
for $x>x'$. The coefficients $g_<$ and $g_>$ are defined in \ref{appa} by Eqs.~(\ref{ci}) and (\ref{ck}), respectively. The
propagator for negative values of the initial condition, i.e., $x'<0$, is obtained from the equations above simply by replacing $x$, $x'$, and $f$ by $-x$, $-x'$, and $-f$, respectively. 

\begin{figure}[t]
\centering
\includegraphics{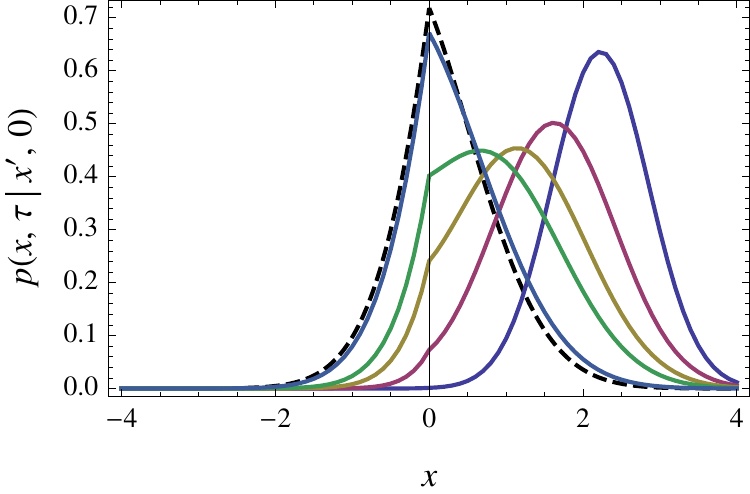}
\caption{(Color online) Propagator $P(x,\tau|x',0)$ in real time obtained by numerically inverting the Laplace transform $\tP(x,s|x',0)$. Parameters: $\delta=1$, $f=0.5$, $x'=3$. The different colored curves moving to the left are obtained for increasing times $\tau=0.25, 0.5, 0.75, 1$ and $2$. The dash curve corresponds to the stationary density $\rho_f(x)$.}
\label{fig1}
\end{figure}

The set of Eqs.~(\ref{bca})-(\ref{bcc}) is the main result of this paper. To gain some insight into the form of this solution, consider the case without dry friction and external forcing, i.e., $\delta=0$ and $f=0$, which corresponds to the Ornstein-Uhlenbeck process. Then $g_<(s,0,0)=1$ and $g_>(s,0,0)=0$, according to Eqs.~(\ref{ci})-(\ref{ck}) of \ref{appa}, and we are led to\footnote{The steps leading to this result yields some interesting integral identities for parabolic cylinder functions, which, to the best of our knowledge, cannot be found in the literature.}
\be
\fl \tP(x,s|x',0) = \frac{\Gamma(s)}{\sqrt{2 \pi}} 
\left\{
\begin{array}{lcl} 
e^{(x')^2/4}\, D_{-s}(x')\, e^{-x^2/4}\, D_{-s}(-x), & & x<x' \\
e^{(x')^2/4}\, D_{-s}(-x')\, e^{-x^2/4}\, D_{-s}(x), & & x>x',
\end{array} \right.
\label{be}
\ee
which is the Laplace transform of the Ornstein-Uhlenbeck propagator. Comparing this particular form of $\tP$ with the general solution above, we see that the products of parabolic cylinder functions occurring in Eqs.~(\ref{bca})-(\ref{bcc}) are essentially the Laplace transform of the Ornstein-Uhlenbeck process, modified to include the dry friction and external forces. Following the method of images, the additional term with the coefficients $g_<$ and $g_>$ can then be considered, at a superficial level, as convolution integrals in the time domain, with $g_<$ and $g_>$ playing the role of the source terms.

From the relatively compact solution of the propagator in Laplace space, we are not able to obtain the propagator itself in closed form, except for special cases, such as the Ornstein-Uhlenbeck process ($\delta=f=0$) and the pure dry friction case corresponding to $\gamma=F=0$. However, for all cases it is possible to obtain the propagator numerically by inverting the Laplace transform. Figure~\ref{fig1} shows the result of this procedure using the so-called Talbot method \cite{talbot1979,abate2004,abate2006}.\footnote{A Mathematica implementation of this method is available at\\ \texttt{http://library.wolfram.com/infocenter/MathSource/5026/}} This figure reproduces exactly our previous results for the propagator based on the spectral decomposition of the Fokker-Planck solution (see Fig.~10 of \cite{touchette2010c}). In general, we have found that the numerical computation of $P(x,\tau|x',0)$ from the inverse Laplace transform of $\tP(x,s|x',0)$ is stable and can be carried out to arbitrary level of accuracy for a large range of physically-relevant parameter values. Our Laplace solution can therefore be considered a useful complement to the spectral solution presented in \cite{touchette2010c}.

\section{Power spectrum}

Using the closed-form solution for $\tP(x,s|x',0)$, we now derive an expression for the Laplace transform of the auto-correlation function:
\be
\tC (s)
=\int_0^\infty e^{-st}\, \langle x(t)x(0)\rangle\, dt.
\ee
In terms of $\tP(x,s|x',0)$, we thus have
\be
\tC(s)=\int_{-\infty}^\infty dx' \int_{-\infty}^\infty dx \, x\,x'\, \tP(x,s|x',0)\, \rho_f(x'),
\ee
where $\rho_f(x)$ is the stationary density of the Fokker-Planck equation, given by
\be
\rho_f(x) = \frac{1}{Z}
\left\{
\begin{array}{lcl} 
e^{-(x+\delta-f)^2/2}\, e^{(\delta-f)^2/2} & \mbox{ if } & x>0 \\
e^{-(x-\delta-f)^2/2)}\, e^{(\delta+f)^2/2} & \mbox{ if } & x<0
\end{array} 
\right.
\label{bf}
\ee
with normalization
\be
Z=\frac{D_{-1} (\delta +f)}{D_0(\delta+f)}+\frac{D_{-1} (\delta -f)}{D_0(\delta-f)}.
\label{bg}
\ee
This expression of $\rho_f(x)$ is invariant with respect to the inversions $x\rightarrow -x$ and $f\rightarrow-f$, as is the propagator. As a result, we can rewrite $\tC(s)$ as 
\be
\tilde{C}(s)= 
\int_0^\infty dx' \int_{-\infty}^\infty dx \, ( x\, x'\, \tP(x,s|x',0) \rho_f(x') + `f \rightarrow -f\textrm{'} ), 
\label{bh}
\ee
where the symbol `$f \rightarrow -f$' indicates the contribution to the kernel obtained by replacing $f$ by $-f$.

We proceed to evaluate the integrals in Eq.~(\ref{bh}). The first one in $x$ can be performed using a known relation for the derivative of parabolic cylinder functions, which can be obtained from the relation shown in Eq.~(\ref{cga}) of \ref{appa}. The result is
\begin{eqnarray}
\fl
\int_0^\infty dx' \int_{-\infty}^\infty dx \,  x\, x'\, \tP(x,s|x',0) \rho_f(x')
= \int_0^\infty dx' \,\left(\frac{(x')^2}{s} - \frac{x'(x'+\delta-f)}{s(s+1)}\right) \rho_f(x') \nonumber \\
+ 
\frac{D_{-s-2}(\delta-f) \Gamma(s)}{Z \sqrt{2 \pi}} \Bigg( -\frac{e^{-(\delta+f)^2/4}}{{e^{-(\delta-f)^2/4}}} g_<(s,\delta,f) D_{-s-2}(\delta+f) \nonumber \\
+ 
g_>(s,\delta,f) D_{-s-2}(\delta-f) + D_{-s-2}(-\delta+f)\Bigg).
\label{bi}
\end{eqnarray}
The second term on the right-hand side of this expression can be simplified using the definitions (\ref{ci}) and (\ref{ck}) as well as the identities (\ref{cga}) and (\ref{cgb}):
\begin{eqnarray}\label{bj}
\fl
D_{-s-2}(\delta-f)\Bigg( -\frac{e^{-(\delta+f)^2/4}}{{e^{-(\delta-f)^2/4}}} g_<(s,\delta,f) D_{-s-2}(\delta+f) \nonumber \\
\qquad + g_>(s,\delta,f) D_{-s-2}(\delta-f)+ D_{-s-2}(-\delta+f)\Bigg)\nonumber \\
\fl\qquad = \frac{2 \delta \sqrt{2 \pi}}{s(s+1) \Gamma(s)} 
\frac{D_{-s-1}(\delta+f)}{D_{-s}(\delta-f) D_{-s-1}(\delta+f) + 
D_{-s}(\delta+f) D_{-s-1}(\delta-f)}.
\end{eqnarray}
By combining this result in Eq.(\ref{bh}), and by performing the integral over $x'$, we then obtain
\begin{eqnarray}
\fl \tilde{C}(s) &=& \frac{1}{s} \langle x^2\rangle_f 
- \frac{1}{s(s+1)} \left(  \langle x^2\rangle_f  +
\delta \langle | x|\rangle_f - f \langle x \rangle_f
\right) \nonumber \\
\fl & & + \frac{2\delta}{s(s+1)} \frac{1}{Z} 
\frac{D_{-s-1}(\delta+f) D_{-s-2}(\delta-f) +
D_{-s-1}(\delta-f) D_{-s-2}(\delta+f)}{D_{-s}(\delta+f) 
D_{-s-1}(\delta-f) + D_{-s}(\delta-f) 
D_{-s-1}(\delta+f)},
\label{bk}
\end{eqnarray}
where $\langle \cdot\rangle_f$ denotes the expected value with respect to the stationary distribution $\rho_f$. Surprisingly, all of the stationary expected values appearing above can be conveniently written in terms of parabolic cylinder functions:
\begin{eqnarray}
\langle x^2 \rangle_f &=& 2 \frac{D_0(\delta+f) D_{-3}(\delta-f) + 
D_0(\delta-f) D_{-3}(\delta+f)}{D_0(\delta+f) D_{-1}(\delta-f) + 
D_0(\delta-f) D_{-1}(\delta+f)}\label{bla}\\
\langle |x| \rangle_f &=&  \frac{D_0(\delta+f) D_{-2}(\delta-f) + 
D_0(\delta-f) D_{-2}(\delta+f)}{D_0(\delta+f) D_{-1}(\delta-f) + 
D_0(\delta-f) D_{-1}(\delta+f)}\label{blb}\\
\langle x \rangle_f &=& \frac{D_0(\delta+f) D_{-2}(\delta-f) - 
D_0(\delta-f) D_{-2}(\delta+f)}{D_0(\delta+f) D_{-1}(\delta-f) + 
D_0(\delta-f) D_{-1}(\delta+f)}\label{blc}.
\label{bl}
\end{eqnarray}

The expression shown in Eq.~(\ref{bk}) can be re-written in other ways to make some of its properties more explicit. Using the identity (\ref{cga}), it is easy to check, in particular, that the stationary expectation values obey the relation
\be\label{blda}
\langle x^2\rangle_f +\delta \langle |x|\rangle_f-f \langle x \rangle_f=1,
\ee 
so that
\begin{eqnarray}
\fl \tilde{C}(s) = \frac{1}{s}\left(\langle x^2\rangle_f -1 
+\frac{2 \delta}{Z} 
\frac{D_{-s-1}(\delta+f) D_{-s-2}(\delta-f) +
D_{-s-1}(\delta-f) D_{-s-2}(\delta+f)}{D_{-s}(\delta+f) 
D_{-s-1}(\delta-f) + D_{-s}(\delta-f) 
D_{-s-1}(\delta+f)} \right) \nonumber \\
\fl + \frac{1}{s+1}\left(
1-\frac{2 \delta}{Z} 
\frac{D_{-s-1}(\delta+f) D_{-s-2}(\delta-f) +
D_{-s-1}(\delta-f) D_{-s-2}(\delta+f)}{D_{-s}(\delta+f) D_{-s-1}(\delta-f) 
+ D_{-s}(\delta-f) 
D_{-s-1}(\delta+f)} \right).
\label{bld}
\end{eqnarray}

This form of the Laplace transform of the correlation function, or resolvent, is quite useful to uncover its analytical structure. Because of Eq.~(\ref{bg}) and the identity (\ref{cga}) (for $\nu=0$), the coefficient of $1/(s+1)$ vanishes at $s=-1$ for $\delta\neq 0$. Thus, the apparent singularity at $s=-1$ is removable. For a similar reason, the coefficient of $1/s$ at $s=0$, which corresponds to the residue at $s=0$, is given by $\langle x\rangle_f^2$. Thus, the long-time limit of the correlation function is given, as expected, by the square of the mean velocity. All the other poles of the resolvent $\tilde{C}(s)$ are determined by the denominator of the expression above, i.e., by the zeros of
\be
D_{-s}(\delta+f) D_{-s-1}(\delta-f) + D_{-s}(\delta-f) D_{-s-1}(\delta+f),
\ee 
which is precisely the characteristic equation derived from the Fokker-Planck operator \cite{touchette2010c}.

Another form of $\tC(s)$ can be obtained by rearranging Eq.~(\ref{bld}) using the aforementioned identities to obtain
\begin{eqnarray}
\tilde{C}(s) &=& 
\frac{\langle x^2 \rangle_f-1}{s} + \frac{1}{s+1} + 
\frac{2 \delta/Z}{s(s+1)^2}\nonumber \\
& & -\frac{4 \delta^2/Z}{s(s+1)^2} 
\left( \frac{D_{-s}(\delta+f)}{D_{-s-1}(\delta+f)} +
\frac{D_{-s}(\delta-f)}{D_{-s-1}(\delta-f)} \right)^{-1}.
\label{ble}
\end{eqnarray}
This expression is better suited for numerical calculations than either Eq.~(\ref{blc}) or (\ref{bld}). The limit $\delta=f=0$ is also clearer at the level of this expression. Noting from Eq.~(\ref{blda}) that $\langle x^2\rangle_f=1$ in this limit, we recover $\tilde{C}(s)=1/(s+1)$, which characterizes the simple exponential decay of the correlation function of the Ornstein-Uhlenbeck process.  Finally, we obtain from Eq.~(\ref{ble}) a simple analytic expression for the power spectrum $p(\omega)=\mbox{Re}\, \tilde{C}(s=i\omega)$, i.e., for the Fourier transform of the auto-correlation function, namely,
\begin{eqnarray}
\fl p(\omega) &=& \frac{1}{1+\omega^2}
-\frac{4 \delta/Z}{(1+\omega^2)^2} \nonumber \\
\fl & & 
-\frac{4 \delta^2/Z}{\omega} \mbox{Im} \left(\frac{1}{(1+i\omega)^2} 
\left( \frac{D_{-i \omega}(\delta+f)}{D_{-i\omega-1}(\delta+f)} +
\frac{D_{-i \omega}(\delta-f)}{D_{-i\omega-1}(\delta-f)} \right)^{-1} 
\right).
\label{bm}
\end{eqnarray}
This result is used in the next section to discuss the stick-slip transition occurring at $f=\delta$.

\section{Stick-slip transition}\label{sec3}

\begin{figure}[t]
\centering
\includegraphics{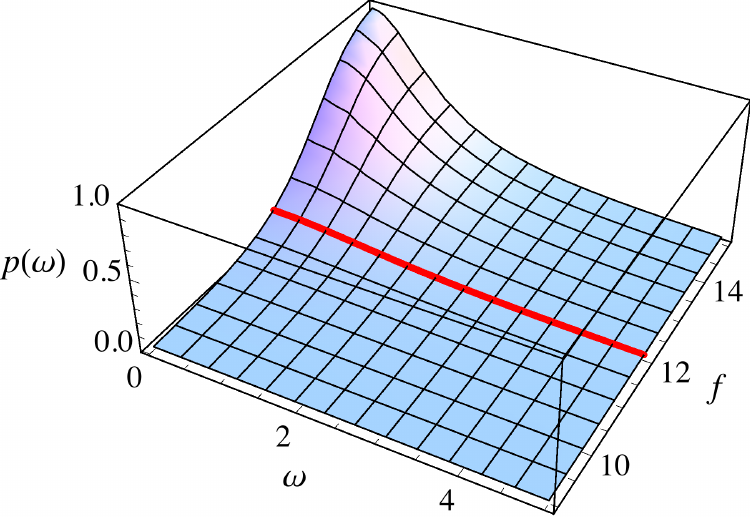}
\caption{(Color online) 3D plot of the power spectrum $p(\omega)$ normalized by the variance $\Delta x^2$ obtained for $\delta=12$ as a function of frequency $\omega$ and the external forcing $f$. The stick-slip transition line $f=\delta$ is shown as the red line.}
\label{fig2}
\end{figure}

We have discussed in detail the behavior of the correlation function $\langle x(t)x(0)\rangle$ as a function of $\delta$ and $f$ in \cite{touchette2010c} and, more precisely, how the stick-slip transition that appears in the deterministic ($\Gamma=0$) equation when $f=\delta$ is modified in the presence of noise ($\Gamma>0$) to a smooth crossover between a stick and a slip regimes, characterized by different exponential decay of $\langle x(t)x(0)\rangle$. Figure~\ref{fig2} shows how this crossover shows up at the level of the power spectrum. We see that $p(\omega)$ is rather flat in the stick regime ($f<\delta$), and that it starts to develop a sharp zero-frequency peak at the stick-slip transition $f=\delta$. This peak is the translation in frequency of the increase of the correlation time associated with $\langle x(t)x(0)\rangle$ as we go from the slip to the stick regimes \cite{touchette2010c}. 

\begin{figure}[t]
\centering
\resizebox{3in}{!}{\includegraphics{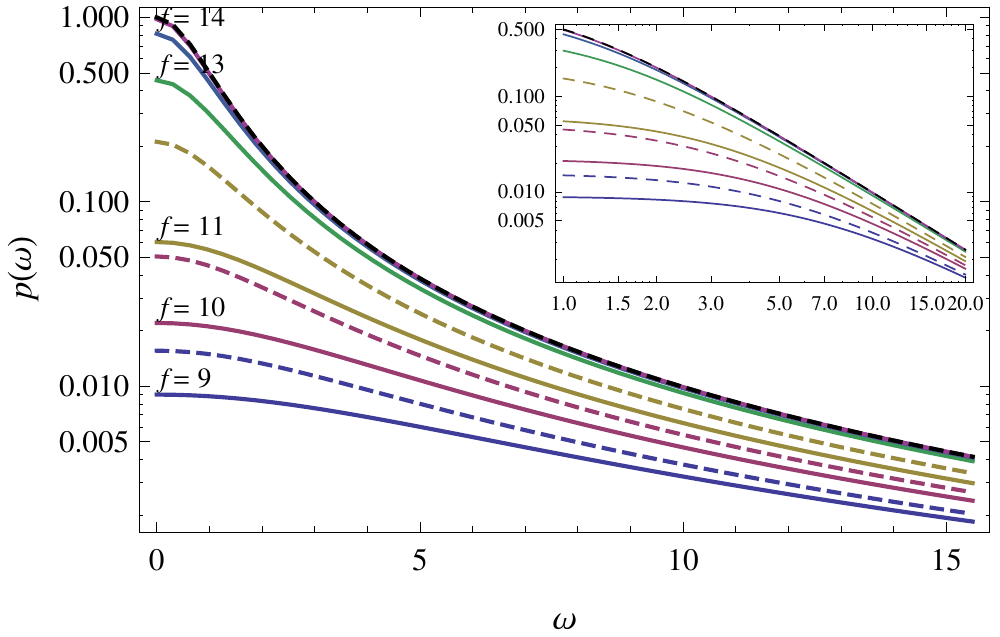}}
\caption{(Color online) Solid lines: Log-linear plot of the power spectrum $p(\omega)$ as function of the frequency $\omega$ for $\delta=12$ and different values of the external forcing $f$ (shown in the plot). Colored dashed lines: Approximation of the power spectrum given by Eq.~(\ref{eb}). Top dashed line in black: Power spectrum of the Ornstein-Uhlenbeck process without dry friction. Inset: log-log plot of $p(\omega)$ showing the $\omega^{-2}$ tail behavior.}
\label{fig3}
\end{figure}

To gain further insight into the behavior of $p(\omega)$, we plot this function on a log-linear scale in Fig.~\ref{fig3} together with an asymptotic expansion of this function obtained in the limit of large dry friction $\delta$ and large forcing $f$. Mathematically, this expansion is equivalent to the small noise limit of the Langevin equation, and is obtained by using the representation (\ref{bld}) for $\tC(s)$ to express the power spectrum as
\begin{eqnarray}
\fl p(\omega) = \frac{1}{\omega} \mbox{Im}\Bigg(\frac{1}{1+i \omega}\nonumber \\
\fl \times
\Big(
\frac{2 \delta}{F_0(\delta+f)+F_0(\delta-f)}\frac{F_{1+i\omega}(\delta+f)
+F_{1+i\omega}(\delta-f)}{
2\delta+(1+i\omega)(F_{1+i\omega}(\delta+f)+F_{1+i\omega}(\delta-f))}-1
\Big)\Bigg),
\label{ea}
\end{eqnarray}
where
\be
F_s(\delta)=\frac{D_{-s-1}(\delta)}{D_{-s}(\delta)}.
\label{eq:ratio}
\ee
With some results of asymptotic analysis, presented in \ref{appb}, we can then obtain the following approximation, which is valid for large values of the parameters $\delta$ and $f$:
\be
\fl p(\omega) \approx
\left\{
\begin{array}{ll}
\frac{1}{\omega} \mbox{Im} \left( 
\frac{\delta^2-f^2}{(1+i \omega)^2} 
\frac{\sqrt{\frac{(\delta+f)^2}{4}+1 
+i\omega}+\sqrt{\frac{(\delta-f)^2}{4}+1 +i\omega}-\delta}
{\sqrt{\frac{(\delta+f)^2}{4}+1 +i\omega}
+\sqrt{\frac{(\delta-f)^2}{4}+1 +i\omega}+\delta}
-\frac{1}{1+i \omega}\right) & 0\leq f <\delta\\
1/(1+\omega^2) \qquad  & 0 \leq \delta < f;
\end{array} 
\right.
\label{eb}
\ee
(see \cite{cuyt2008} for related results derived via continued fraction expansions). Figure~\ref{fig3} shows that this asymptotic formula is relatively accurate, even for rather small parameter values. Larger deviations are visible close to the transition point $\delta=f$. The inset of Fig.~\ref{fig3} also shows that the tail of $p(\omega)$ decays at large frequencies as $\omega^{-2}$, which is the sign that temporal correlations decay exponentially, as in the Ornstein-Uhlenbeck process. 

\begin{figure*}[t]
\centering
\resizebox{\textwidth}{!}{\includegraphics{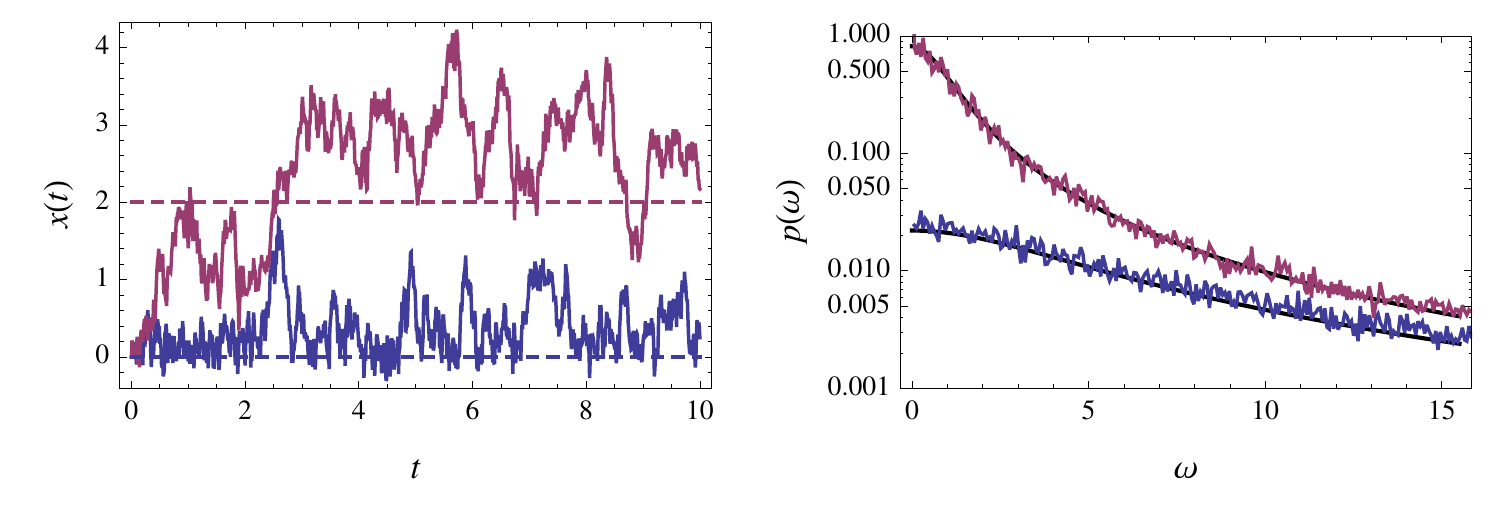}}
\caption{(Color online) Left: Simulated paths of Brownian motion with dry friction (for the rescaled variable $x$) for $\delta=12$ and $f=10$ (blue, stick regime) and $f=14$ (purple, slip regime). The numerical integration was done with the Euler-Maruyama method with $\Delta t=0.01$. The steady-state velocity for the slip regime is $x^*=f-\delta=2$. Right: Corresponding power spectra (colored lines) compared with the theory (black line). The numerical spectra were obtained by averaging 50 times series over the time interval $[0,100]$.}
\label{fig4}
\end{figure*}

To give an idea of how these results might compare in practice with experimental results, we show in Fig.~\ref{fig4} simulated paths of the Brownian motion equation with dry friction for the rescaled variable $x(t)$ together with their corresponding power spectrum. Two paths are shown on the left-hand side of Fig.~\ref{fig4}: one in the stick regime ($\delta=12$ and $f=10$) and one in the slip regime ($\delta=12$ and $f=14$). The power spectrum characterizing each of these regimes, shown on the right-hand side of Fig.~\ref{fig4}, is computed numerically by averaging the spectra of many random paths (here 50) over a relatively long time (here $T=100$). The numerical results compare well with the theory, as can be seen. Instead of averaging different spectra, one can also use, as is well known, a frequency window larger than the frequency spacing $\Delta\omega=2\pi/T$ to obtain a relatively smooth spectrum.

\begin{figure}[t]
\centering
\resizebox{\textwidth}{!}{\includegraphics{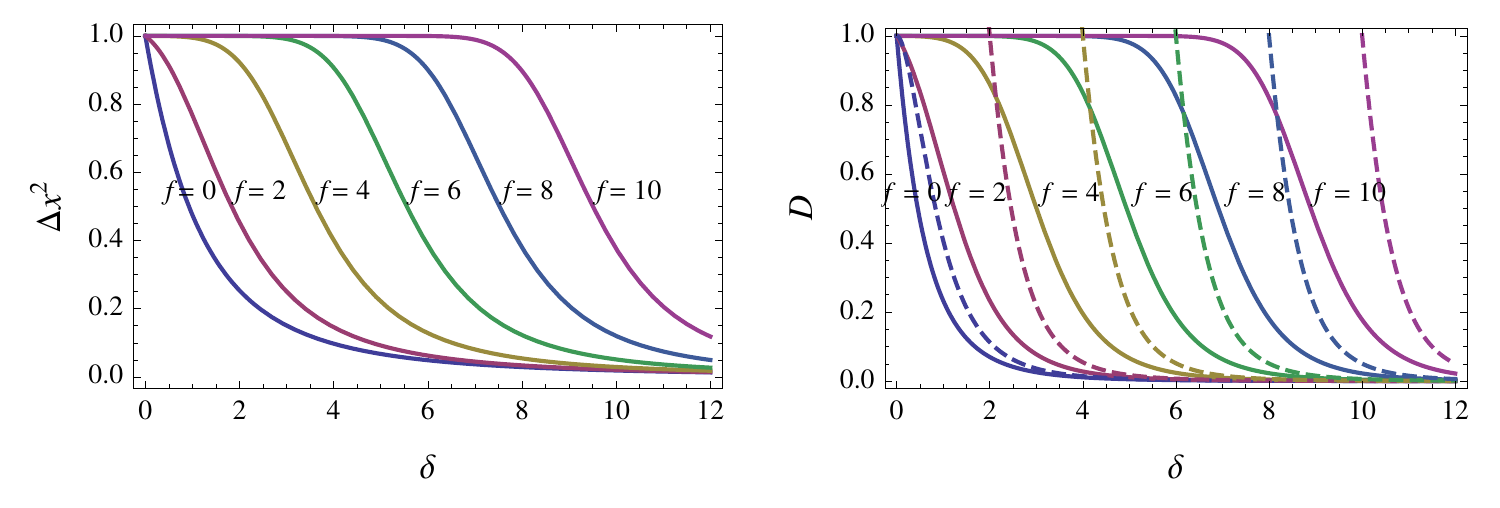}}
\caption{(Color online) Left: Covariance $\Delta x^2$ given by Eqs.~(\ref{bla}) and (\ref{blc}) as a function of the dry friction and driving forces. Right: Diffusion constant $D=p(\omega\rightarrow 0)$ as a function of $\delta$ for various values of $f$. The solid line is the exact expression shown in Eq.~(\ref{ea}), whereas the dashed line is the diffusion constant obtained from the approximation shown in Eq.~(\ref{eb}).}
\label{fig5}
\end{figure}

To close this section, let us now briefly discuss the behavior of two quantities derived from the power spectrum. The first is the stationary variance $\Delta x^2$, which is proportional to the total spectral weight:
\be
\Delta x^2=\langle x^2\rangle_f-\langle x \rangle_f^2=\frac{1}{\pi}\int_{-\infty}^\infty p(\omega) d \omega.
\label{eqsv1}
\ee
As shown in Fig.~\ref{fig5}, there is a sharp increase of $\Delta x^2$ at the stick-slip transition $f=\delta$, separating a low variance (stick) regime from a high variance (slip) regime.  The second quantity is the (non-dimensional) diffusion constant:
\be
D=\int_0^\infty \langle x(t)x(0)\rangle\, dt = 
\lim_{\omega\rightarrow 0}\, p(\omega).
\ee
This quantity is of particular interest, since it can be measured in experiments. Its expression is obtained from Eqs.~(\ref{ea}) or (\ref{eb}) and is plotted in Fig.~\ref{fig5}. We see that in the slip regime, the diffusion constant does not depend much on the external force, as is the case for the Ornstein-Uhlenbeck process. This is consistent with the observation that the slip regime is essentially a regime of normal Brownian motion in which dry friction force plays little role; see \cite{touchette2010c} for more details. In the stick regime, on the other hand, the diffusion constant is relatively small, and sharply increases when the stick-slip transition is approached. This behavior is also seen if we derive $D$ from the asymptotic expression (\ref{eb}). This is illustrated with the dashed lines in Fig.~\ref{fig5}.

\section{Conclusion}
\label{sec4}

We have presented the exact solution of the propagator of a Langevin equation modeling Brownian motion in the presence of solid friction, viscous damping, and an external constant force. The main feature of this equation is that it shows a stick-slip transition often encountered in real systems involving solid friction and external forcing. The solution of this equation follows by considering the Laplace transform of its associated Fokker-Planck equation, and serves as a complement to a previous exact solution, derived in \cite{touchette2010c} using the spectral decomposition of the Fokker-Planck operator. It also extends a similar Laplace solution, previously derived in \cite{caughey1961,atkinson1968,atkinson1968b} for the special case where only solid friction is present.

A clear advantage of the Laplace solution over the spectral solution is that the former is given as an explicit and compact formula, which can be used to obtain the propagator by inverse Laplace transform. The Laplace solution is also useful as it enables us to obtain the power spectrum of the system, in addition to the diffusion constant of the Brownian motion affected by solid friction. Both of these characteristics are easily accessible experimentally, and so might be useful to compare the model with experiments involving noise and solid friction, such as those recently reported for example in \cite{goohpattader2009,goohpattader2010}. 

\section*{Acknowledgments}

H.T.\ is grateful for the support and hospitality of the National Institute of Theoretical Physics at the University of Stellenbosch, South Africa, where part of this work was written. The work of W.J.\ is partly supported by EPSRC through grant no.\ EP/H04812X/1.

\appendix
\section{Laplace transform of the propagator}
\label{appa}

We provide here some details of the derivation of the expressions (\ref{bca})-(\ref{bcc}) for the Laplace transform $\tP(x,s|x',0)$ of the propagator $P(x,t|x',0)$.

The derivation starts with the Fokker-Planck equation (\ref{ba}), which for positive values for the initial condition $x'>0$ reads
\begin{eqnarray}
\fl s \tP(x,s|x',0) &=& \frac{d}{d x} (x-\delta-f) \tP(x,s|x',0)+ \frac{d^2 \tP}{d x^2}, \qquad x<0\label{caa}\\
\fl s \tP(x,s|x',0) &=& \frac{d}{d x} (x+\delta -f) \tP(x,s|x',0)+ \frac{d^2 \tP}{d x^2},\qquad 0<x<x'\label{cab}\\
\fl s \tP(x,s|x',0) &=& \frac{d}{d x} (x+\delta -f) \tP(x,s|x',0)+ \frac{d^2 \tP}{d x^2}, \qquad x'<x.\label{cac}
\end{eqnarray}
These equations must be solved with the following matching conditions at $x=0$:
\begin{eqnarray}
\hspace*{0.25in}\tP(x=0^-,s|x',0) &=& \tP(x=0^+,s|x',0) \label{cba}\\
\fl(-\delta-f) \tP(x=0^-,s|x',0) + \left.\frac{d \tP}{dx}\right|_{x=0^-} &=& (\delta-f) \tP(x=0^+,s|x',0) + \left.\frac{d \tP}{dx}\right|_{x=0^+}\label{cbb}\\
\hspace*{0.07in}\tP(x=x'-0,s|x',0) &=& \tP(x=x'+0,s|x',0) \label{cbc}\\
\hspace*{1.15in}-1 &=& \left.\frac{d \tP}{dx}\right|_{x=x'+0}-\left.\frac{d \tP}{dx}\right|_{x=x'-0}\label{cbd},
\end{eqnarray}
which results from the continuity of the probability current, as well as the usual decaying boundary conditions at $x=\pm\infty$. 

The equations (\ref{caa})-(\ref{cac}) have the form of the Hermite differential equation 
\be
u''(z)+(zu(z))'+\nu u(z)=0.
\label{cc}
\ee
The solution can be written in terms of parabolic cylinder functions either as $e^{-z^2/4} D_\nu(z)$ or $e^{-z^2/4} D_\nu(-z)$. \footnote{These two solutions are linearly independent, i.e., they constitute a fundamental system if the index $\nu$ is not an integer. For a complete account
of parabolic cylinder functions, the reader may consult \cite{buchholz1969}.}.
In view of the asymptotic property of the parabolic cylinder function,
\be
D_\nu(z) \sim z^\nu\, e^{-z^2/4},\qquad z\rightarrow \infty,
\ee
the solution of Eqs.~(\ref{caa}) and (\ref{cac}) having vanishing currents at infinity can be written as
\be
\tP(x,s|x',0) = C_-(x',s)\, e^{-(x-\delta-f)^2/4}\, D_{-s}(-x+\delta+f)
\label{cda}
\ee
for $x<0$ and
\be
\tP(x,s|x',0) =C_+(x',s) \, e^{-(x+\delta-f)^2/4}\, D_{-s}(x+\delta-f)
\label{cd}
\ee
for $x>x'$. On the other hand, the solution of Eq.~(\ref{cab}) is given by a linear combination of the two fundamental solutions:
\begin{eqnarray}\label{ce}
\fl \tP(x,s|x',0) &=& B_+(x',s)\, e^{-(x+\delta-f)^2/4}\, D_{-s}(x+\delta-f) \nonumber \\
\fl & & + B_{-}(s,x')\, e^{-(x+\delta-f)^2/4}\, D_{-s}(-x-\delta+f), 
\qquad 0<x<x'.
\end{eqnarray}
The amplitudes $B_\pm$ and $C_\pm$ are determined by the matching conditions (\ref{cba})-(\ref{cbd}), which result in a set of inhomogeneous linear equations:
\begin{eqnarray}
\fl \hspace*{0.25in} C_-(x',s)\, e^{-(\delta+f)^2/4}\, D_{-s}(\delta+f) &=& B_+(x',s)\, e^{-(\delta-f)^2/4}\, D_{-s}(\delta-f) \nonumber \\
\fl & & + B_{-}(s,x')\, e^{-(\delta-f)^2/4}\, D_{-s}(-\delta+f) \label{cfa}\\
\fl -C_-(x',s)\, e^{-(\delta+f)^2/4}\, D_{-s-1}(\delta+f) &=& B_+(x',s)\, e^{-(\delta-f)^2/4}\, D_{-s-1}(\delta-f) \nonumber \\
\fl & & - B_-(x',s)\, e^{-(\delta-f)^2/4}\, D_{-s-1}(-\delta+f) \label{cfb}\\
\fl \hspace*{0.63in} C_+(x',s) D_{-s}(x'+\delta-f) &=& B_+(x',s) D_{-s}(x'+\delta-f) \nonumber \\
\fl & & + B_-(x',s) D_{-s}(-x'-\delta+f) \label{cfc}\\
\fl \hspace*{1.4in} e^{(x'+\delta-f)^2/4} &=& C_+(x',s) D_{-s+1}(x'+\delta-f)\nonumber\\
\fl & &- B_+(x',s) D_{-s+1}(x'+\delta-f) \nonumber \\
\fl & & + B_-(x',s) D_{-s+1}(-x'-\delta+f). \label{cfd}
\label{cf}
\end{eqnarray}
In writing these equations, we have used the following identities for the parabolic cylinder functions:
\begin{eqnarray}\label{cga}
\nu\, e^{-z^2/4}\, D_{\nu-1}(z) &=& e^{-z^2/4} (z D_\nu(z)-D_{\nu+1}(z)) \nonumber \\
&=& z\,  e^{-z^2/4}\, D_\nu(z) +(e^{-z^2/4}\, D_\nu(z))'
\end{eqnarray}
to evaluate and simplify the derivatives. 

Equations~(\ref{cfa})-(\ref{cfd}) can be solved directly to find $B_{\pm}$ and $C_{\pm}$. The difference of Eqs.~(\ref{cfc}) and (\ref{cfd}) yields
\be
B_-(x',s) = \frac{\Gamma(s)}{2 \sqrt{\pi}}\, e^{(x'+\delta-f)^2/4}\, D_{-s}(x'+\delta-f),
\label{cg}
\ee
if we take into account the product identity 
\be
D_\nu(z) D_{\nu-1}(-z)+D_\nu(-z) D_{\nu-1}(z)=\frac{\sqrt{2\pi}}{\Gamma(-\nu+1)},
\label{cgb}
\ee
which follows from the Wronskian of the fundamental system.\footnote{Equation~(\ref{cgb}) corrects a typo in Eq.~(33a) of \cite{buchholz1969}.} Then the difference of Eqs.~(\ref{cfa}) and (\ref{cfb}) yields
\be
C_-(x',s)=g_<(s,\delta,f) B_-(x',s),
\label{ch}
\ee
where
\be
 g_<(s,\delta,f) = \frac{\displaystyle 
e^{-(\delta-f)^2/4} \left(\frac{D_{-s}(-\delta+f)}{D_{-s}(\delta-f)}+
\frac{D_{-s-1}(-\delta+f)}{D_{-s-1}(\delta-f)}\right)}{\displaystyle 
e^{-(\delta+f)^2/4}
\left(\frac{D_{-s}(\delta+f)}{D_{-s}(\delta-f)}+
\frac{D_{-s-1}(\delta+f)}{D_{-s-1}(\delta-f)}
\right)}.
\label{ci}
\ee
Summing Eqs.~(\ref{cfa}) and (\ref{cfb}), we also find
\be
B_+(x',s)=g_>(s,\delta,f) B_-(x',s)
\label{cj}
\ee
with
\be
g_>(s,\delta,f)=\frac{\displaystyle 
\frac{D_{-s-1}(-\delta+f)}{D_{-s-1}(\delta+f)}
-\frac{D_{-s}(-\delta+f)}{D_{-s}(\delta+f)}}{\displaystyle 
\frac{D_{-s}(\delta-f)}{D_{-s}(\delta+f)}+
\frac{D_{-s-1}(\delta-f)}{D_{-s-1}(\delta+f)}}.
\label{ck}
\ee
Finally, summing Eqs.~(\ref{cfc}) and (\ref{cfd}), we find 
\be
C_+(x',s)=B_+(x',s)+\frac{\Gamma(s)}{\sqrt{2\pi}}\, e^{(x'+\delta-f)^2/4}\, D_{-s}(-x'-\delta+f).
\label{cl}
\ee
The substitution of all these expressions back into Eqs.~(\ref{cda})-(\ref{ce}) leads us to Eqs.~(\ref{bca})-(\ref{bcc}).

\section{Asymptotic expansion}
\label{appb}

The power spectrum, defined in Eq.~(\ref{bm}), is entirely determined by the ratio $F_s(\delta)$ of parabolic cylinder functions defined in Eq.~(\ref{eq:ratio}). Using the linear recurrence relation (\ref{cga}), this ratio is found to obey the relation
\be
F_s(\delta)=\frac{1}{\delta + (s+1) F_{s+1}(\delta)},
\ee
iteration of which leads to continued fraction expansions \cite{cuyt2008}.

Asymptotic expressions for $F_s(\delta)$ for large values of $\delta$ and $s$ can be obtained from the contour integral representation of parabolic cylinder functions,
\be
D_\nu(z)=\frac{e^{-z^2/4}}{i\sqrt{2\pi}}\int_{c-i\infty}^{c+i\infty}e^{-zt+t^2/2}t^\nu dt,
\ee
which is valid for all $z,\nu\in\mathbb C$, and any integral contour satisfying $|\arg t|<\pi/2$ and $c>0$. To this end, we introduce a large parameter $N$ and change the variable of integration to $u=t/\sqrt{N}$. This leads to
\be
F_{Ns}(\sqrt{N}\delta)=\frac1{\sqrt{N}} \,
\frac{\displaystyle\int_{c-i\infty}^{c+i\infty}e^{N(-\delta u+u^2/2-s\log u)}\frac{du}u}
{\displaystyle\int_{c-i\infty}^{c+i\infty}e^{N(-\delta u+u^2/2-s\log u)}du}.
\ee
Each of the integrals above has the Laplace form
\be
\int_{\cal C}e^{Nf(z)}g(z)dz,
\ee
which can be approximated following the saddlepoint method as
\be
\int_{\cal C}e^{Nf(z)}g(z)dz\sim \pm i\, e^{Nf(z_s)}g(z_s)\sqrt{\frac{2\pi}{N f''(z_s)}},
\ee
where $z_s$ is the saddlepoint of the integral given by $f'(z_s)=0$, and the sign depends on the orientation of the contour $C$, which is deformed to a steepest-descent contour passing through the saddlepoint. This approximation is valid to leading order in $N$ provided $g$ does not vanish at $z_s$.

In our case, we obtain the saddlepoints from
\be
0=\frac\p{\p u}\left(-\delta u+\frac12u^2-s\log u\right)=-\delta+u-\frac su,
\ee
which has two solutions:
\be
u_1=\frac12\left(\delta+\sqrt{\delta^2+4s}\right),\qquad
u_2=\frac12\left(\delta-\sqrt{\delta^2+4s}\right).
\ee
On the one hand, if $\delta^2+4s>0$, then $u_1>0>u_2$ and the steepest-descent contour passes through $u_1$ only. In this case, much of the saddlepoint approximation cancels, leaving us with
\be
F_{Ns}(\sqrt{N}\delta)\sim\frac1{\sqrt{N}}\frac1{u_1}=\frac2{\sqrt{N}\delta+\sqrt{N\delta^2+4Ns}}.
\label{db}
\ee
On the other hand, if $\delta^2+4s<0$, then $u_1$ and $u_2$ are complex conjugate to each other, and the steepest-descent contour passes through both saddlepoints. The crossover between these cases is given by coalescing saddlepoints; a uniform asymptotic expansion for this case can be obtained in terms of Airy functions \cite{wong1989}.

A second case of interest is given when the argument of $F$ is kept constant, i.e.,
\be
F_s(\sqrt{N}\delta)=\frac1{\sqrt{N}}\,
\frac{\displaystyle\int_{c-i\infty}^{c+i\infty}e^{N(-\delta u+u^2/2)}\frac{du}{u^{s+1}}}
{\displaystyle\int_{c-i\infty}^{c+i\infty}e^{N(-\delta u+u^2/2)}\frac{du}{u^s}}.
\ee
In this case, there is only one saddlepoint at $u_0=\delta$, which collides with a singularity of the integrand when $\delta$ approaches zero. For $\delta>0$, we then find
\be
F_s(\sqrt{N}\delta)\sim\frac1{\sqrt{N} u_0}=\frac1{\sqrt{N}\delta},
\ee
which matches the previous asymptotic evaluation. For $\delta<0$, i.e., for Kramer's problem, the integral is dominated by the singularity at the origin, and its contribution gives
\be
F_s(\sqrt{N}\delta)\sim\frac{-\sqrt{N}\delta}s
\ee
for $s\neq0,-1,-2,\ldots$. If $s=-1,-2,\ldots$, we find instead 
$F_s(\sqrt{N}\delta)\sim1/(\sqrt{N}\delta)$.

\section*{References}
\bibliographystyle{unsrt}
\bibliography{dflaplace} 

\end{document}